\documentstyle[epsfig]{mn}
\begin{document}

\def\di{\mbox{d}}
\def\Msun{M_{\odot}}
\def\HI{\hbox{H$\scriptstyle\rm I\ $}}
\def\HII{\hbox{H$\scriptstyle\rm II\ $}}
\def\mpc{\,{\rm {Mpc}}}
\def\kpc{\,{\rm {kpc}}}
\def\kms{\,{\rm {km\, s^{-1}}}}
\def\msun{{M_\odot}}
\def\Gyr{{\,\rm Gyr}}
\def\erg{{\rm erg}}
\def\sr{{\rm sr}}
\def\hz{{\rm Hz}}
\def\cm{{\rm cm}}
\def\sec{{\rm s}}
\def\eV{{\rm \ eV}}

\def\gtsima{$\; \buildrel > \over \sim \;$}
\def\ltsima{$\; \buildrel < \over \sim \;$}
\def\prosima{$\; \buildrel \propto \over \sim \;$}
\def\gsim{\lower.5ex\hbox{\gtsima}}
\def\lsim{\lower.5ex\hbox{\ltsima}}
\def\simgt{\lower.5ex\hbox{\gtsima}}
\def\simlt{\lower.5ex\hbox{\ltsima}}
\def\simpr{\lower.5ex\hbox{\prosima}}
\def\la{\lsim}
\def\ga{\gsim}

\def\ie{{\frenchspacing\it i.e. }}
\def\eg{{\frenchspacing\it e.g. }}

\newcommand{\q}{\begin{equation}}
\newcommand{\qa}{\begin{eqnarray}}
\newcommand{\qs}{\begin{eqnarray*}}
\newcommand{\nq}{\end{equation}}
\newcommand{\nqa}{\end{eqnarray}}
\newcommand{\nqs}{\end{eqnarray*}}

\title[Reionization History]
{Reionization History from Coupled CMB/21cm Line Data}            

\author[Salvaterra, Ciardi, Ferrara, Baccigalupi]
{R. Salvaterra$^1$, B. Ciardi$^2$, A. Ferrara$^1$, C. Baccigalupi$^1$\\
$^1$ SISSA/International School for Advanced Studies, Via Beirut 4, 34014 Trieste, Italy\\
$^2$ Max-Planck-Institut f\"{u}r Astrophysik, Garching, Germany}

\maketitle \vspace {7cm }

\begin{abstract}
We study CMB secondary anisotropies produced by inhomogeneous reionization 
by means of cosmological simulations coupled with the radiative transfer code 
{\tt CRASH}. The reionization history is consistent with the WMAP Thomson optical 
depth determination. We find that the signal arising from this process dominates
over the primary CMB component for $l\gsim 4000$ and reaches a 
maximum amplitude of $l(l+1)C_l/2\pi\simeq 1.6\times 10^{-13}$ on arcmin scale,
i.e. $l$ as large as several thousands. 
We then cross-correlate secondary CMB anisotropy maps with 
neutral hydrogen 21cm line emission fluctuations obtained from the same simulations.
The two signals are highly anti-correlated on angular scales corresponding 
to the typical size of \HII regions (including overlapping) at the 21cm map redshift. 
We show how the CMB/21cm cross-correlation can be used to: 
(a) study the nature of the reionization sources, (b)  
reconstruct the cosmic reionization history, (c) infer the mean cosmic     
ionization level at any redshift. We discuss the feasibility of the proposed
experiment with forthcoming facilities. 
\end{abstract}


\section{Introduction}
The Wilkinson Microwave Anisotropy Probe (WMAP\footnote{http://map.gsfc.nasa.gov})
has provided strong evidence for an optical depth to Thomson scattering of
$\tau_e\sim 0.17\pm0.04$ (the uncertainty quoted for this number depends on the
analysis technique employed), based on the measured correlation between Cosmic Microwave
Background (CMB) temperature and polarization on large angular scales
(e.g. Kogut et al. 2003). If the reionization process is
described as instantaneous and homogeneous, this corresponds to a reionization
redshift $z_{ion} \sim 17\pm 5$. More probably, reionization went through a highly
inhomogeneous phase (e.g. Ciardi et al. 2000; Gnedin 2000; Miralda-Escud\'e,
Haehnelt \& Rees 2000; Ciardi, Stoehr \& White 2003, CSW; Sokasian et al. 2003; Ricotti \&
Ostriker 2004),
which ended only when the individual \HII regions overlapped completely.
In this case, the reionization process should have left an imprint on the CMB.
In fact, the modulation of the ionization fraction, playing a similar role to the
density modulation from the non-linear Vishniac effect, leads to anisotropies at
sub-degree scales (e.g. Bruscoli et al. 2000;  Benson et al. 2001; Gnedin \& Jaffe 2001;
Santos et al. 2003).
In addition to temperature anisotropies, Thomson scattering introduces a polarization
signal in the CMB spectrum.
The detection of anisotropies in the temperature/polarization power spectrum is
an invaluable tool to discriminate between different sources of ionizing photons and
reionization histories (e.g. Bruscoli, Ferrara \& Scannapieco 2002; Holder et al. 2003;
Naselsky \& Chiang 2004).

An alternative way to probe the end of the cosmic `dark ages' is through 21cm tomography.
From the pioneering work of Field (1959), it has been suggested that
the neutral hydrogen in the Intergalactic Medium (IGM)
and in gravitationally collapsed systems may be
detectable in emission or absorption against the CMB at the frequency corresponding
to the redshifted 21cm line associated with the spin-flip transition of the hyperfine 
levels of neutral hydrogen. The inhomogeneities in the density field, ionized hydrogen
and spin temperature produce signatures both in the angular and in the redshift
space. Different signatures have been investigated, ranging from the 21cm line emission
induced by the `cosmic web' (Madau, Meiksin \& Rees 1997; Tozzi et al. 2000), the
neutral hydrogen surviving reionization (e.g. Ciardi \& Madau 2003; Furlanetto, Sokasian \&
Hernquist 2004; Furlanetto, Zaldarriaga \& Hernquist 2004) or the minihalos with virial
temperatures below $10^4$~K (e.g. Iliev et al. 2002), to the 21cm lines generated in
absorption against very high-redshift radio sources by the neutral IGM (Carilli, Gnedin
\& Owen 2002) and by intervening minihalos and protogalactic disks (Furlanetto \& Loeb 2002).

In this paper, we compute the CMB temperature anisotropies due to
an inhomogeneous reionization history obtained from radiative transfer simulations
consistent with WMAP observations (Ciardi, Ferrara \& White 2003, hereafter CFW).
Moreover, we cross-correlate them with the expected 21cm
emission maps obtained by Ciardi \& Madau (2003, hereafter CM) for the same
simulations, and discuss how the cross-correlation can be used to reconstruct
the reionization history and to constrain the nature of ionizing sources.
Our work is similar in spirit to the recently published study by Cooray
(2004), although that work is based on a simplified analytical description 
of the reionization process. This might be the reason for which our
conclusions differ from those obtained by Cooray (see Section~\ref{sec:concl}). 

The paper is organized as follows. In Section~\ref{sec:num} we present the
numerical simulations of IGM reionization by CSW and CFW,
and in Section~\ref{sec_21cm} the results of CM on the 21cm emission from
such patchy reionization histories are briefly described. In Section~\ref{sec:CMB} we construct and study the maps and the angular power spectra for 
secondary CMB temperature anisotropies due to the above 
reionization process, whereas the cross-correlation between CMB and
21cm maps is presented  in Section~\ref{sec:cc}. Finally, in Section~\ref{sec:concl} we summarize and discuss the results.

Throughout the paper we adopt the $\Lambda$CDM ``concordance'' model
with $\Omega_m$=0.3, $\Omega_{\Lambda}$=0.7, $h=$0.7, $\Omega_b$=0.04,
$n$=1 and $\sigma_8$=0.9, within the WMAP experimental error bars
(Spergel et al. 2003).

\section{Numerical Simulations of IGM Reionization}
\label{sec:num}

In this Section we briefly describe the numerical simulations of IGM
reionization adopted to model the 21cm line emission from neutral
IGM and the CMB temperature anisotropies, and refer to CSW and CWF 
for further details.

A cosmological volume of comoving side 479$h^{-1}$~Mpc has been simulated
(Yoshida, Sheth \& Diaferio 2001) with the N-body code {\tt GADGET} (Springel,
Yoshida \& White 2001). An approximately spherical region with a diameter
of about $50 h^{-1}$~Mpc has been subsequently ``re-simulated'' at a higher
resolution (Stoehr 2003) with the technique described in Springel et al. (2001,
hereafter SWTK). A friends-of-friends algorithm was employed to determine
the location and mass of dark matter halos.
Gravitationally bound substructures
have been identified within the halos with the algorithm {\tt SUBFIND}
(SWTK) and have been used to build the merging tree for halos and subhalos
following the prescription of SWTK. A particle mass of $M_p=1.7
\times 10^8 h^{-1}$~M$_\odot$ allows to resolve halos as small as
$M \simeq 10^9$~M$_\odot$. The galaxy population has been modeled with
the semi-analytic technique described in Kauffmann et al. (1999) and 
implemented as in SWTK. For each of the simulation output we compile
a catalogue of galaxies containing for each galaxy, among
other quantities, its position, mass and star formation rate. 

A cube of comoving side $L=20 h^{-1}$~Mpc has been cut from the high resolution
spherical subregion to model the details of the reionization process,
using the radiative transfer code {\tt CRASH} (Ciardi et al. 2001;
Maselli, Ferrara \& Ciardi 2003) to follow the propagation into the IGM
of the ionizing photons emitted by the simulated galaxy population.
Several sets of radiative transfer simulations have been run in CSW and
CFW, with different
choices for the galaxy emission properties. The ones used here are
labeled S5 (`late' reionization case) and L20 (`early' reionization
case), and adopt an emission spectrum typical of Pop~III stars, a
Salpeter Initial Mass Function (IMF) and an escape fraction of
ionizing photons $f_{esc}=5\%$ (S5) and a Larson IMF with $f_{esc}=20\%$
(L20). For details and discussion on the choice of parameters
we refer to CSW and CFW. The S5 and L20 simulations give a reionization redshift of
$z_{ion}\approx 8$ and $\approx 13$, respectively. In addition, they provide the redshift
evolution and the spatial distribution of ionized/neutral IGM and have been
used to model both the \HI 21cm line emission (see Sec.~\ref{sec_21cm}) and the
CMB temperature anisotropies (see Sec.~\ref{sec:CMB}).

\section{Secondary CMB Anisotropies}
\label{sec:CMB}

The solution of the Boltzmann equation for the present value of the 
perturbation of the photon temperature $(\delta T/T)_{\rm CMB}$, 
along the line of sight (los) $\hat\gamma$, can be written as:

\q\label{eq:dt/t}
\left( \frac{\delta T}{T}\right)_{\rm CMB}(\hat\gamma) =\tau_0 \int_{0}^1 {\frac{d\eta }{\eta^4}} \chi({\bf x}%
,\eta) \hat\gamma \cdot {\bf v}({\bf x},\eta),
\nq

\noindent
where $\eta=\int (1+z) dt$ is the conformal time and
$\tau_0=n_{e,0}\sigma_T\eta_0 c$, with $n_{e,0}$ being the present free electron
density, $\sigma_T$ the Thomson cross section and $c$ the light speed. The quantities 
$\chi({\bf x},\eta)$ and ${\bf v}({\bf x},\eta)$ are the ionization fraction 
and the peculiar velocity in units of $c$, respectively, 
calculated at position ${\bf x}=\hat\gamma(\eta_0-\eta)$ and conformal time
$\eta$.

In principle, a map of temperature anisotropies can be simply obtained by integrating 
eq.~(\ref{eq:dt/t}) along each los passing trough 
random slices of the simulation boxes. However, the periodic 
simulation boundary conditions would artificially enhance the anisotropy
signal by a non-negligible factor (Gnedin \& Jaffe 2001). To prevent this spurious effect,
we randomly flip and transpose each simulation box around any of its 
six edges, hence breaking the fictitious correlations introduced by 
the computational method. We consider 30 (65) simulation outputs from $z=18.7$ 
to complete reionization, i.e. $z \approx 13$ ($z \approx 8$) for the L20 (S5) model.
The output redshifts are optimized to completely cover the path between the initial 
and final redshift of the simulation.
Although this method might now somewhat underestimate
the true anisotropy signal as we miss the contribution of scales larger
than the box, the results constitute a solid lower limit to such quantity.
In addition, we emphasize that the size of our box ($L=20 h^{-1}$~Mpc) is one
of the largest used up to now for reionization studies, hence making the 
large-scale missing power a less severe effect.

The dimension of the map is set
by the angle subtended by the simulation box at the highest 
redshift; for the adopted cosmology
$\theta_{max}\approx 9.25$ arcmin.  The spatial information on the 
ionization fraction is obtained from the radiative transfer simulations, 
whereas the peculiar velocity field is provided by the $N$-body simulation.
We repeat the above integration for a random realization with the same
volume-averaged value of the ionization fraction $\chi$.
A map of the temperature fluctuations due to ionized patches (i.e.
the inhomogeneous part) is derived by subtracting the two maps. The
result is shown in Fig.~\ref{fig:maps} for the S5 (top left panel)
and the L20 (top right panel) model. 

\subsection{Anisotropy distribution}

The statistics of temperature anisotropy can be analyzed in terms of spherical 
harmonics, $Y_{lm}$:

\q
\left( \frac{\delta T}{T}\right)_{\rm CMB}(\hat\gamma)=
\sum_{l=1}^{\infty}\sum_{m=-l}^{l} a_{lm} Y_{lm}(\hat\gamma),
\nq

\noindent
The angular power spectrum, $C_l$, is then defined as:

\q
C_l\equiv <|a_{lm}^2|>=\frac{1}{2l+1}\sum_m a_{lm}^2.
\nq

\begin{figure}
\center{{
\epsfig{figure=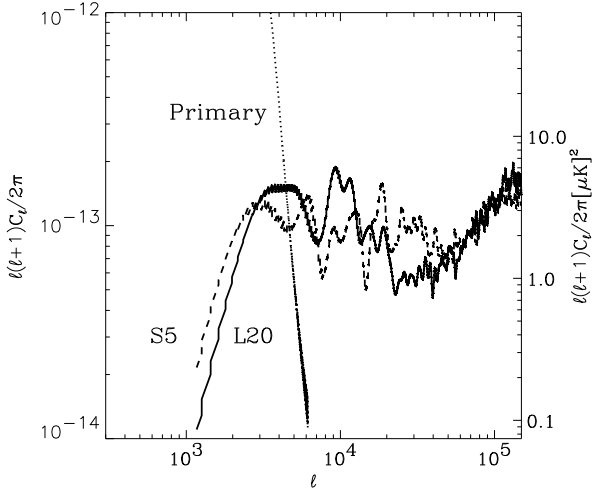,width=8.5cm}
}}
\caption{Primary CMB power spectrum from WMAP data fitting (dotted line; 
Spergel et al. 2003),
together with the angular power spectra from patchy reionization in the L20
(solid) and S5 (dashed) model.}
\label{fig:power}
\end{figure}

There is a strict relation between the probed angular scale $\theta$ and the multipole 
$l$ in the formula above, $\theta\simeq 180/l$ degrees. Therefore the extension 
and resolution of our maps sets an interval in $l$ in which our analysis is 
meaningful, i.e. $4000\le l\le 1.67\cdot 10^{5}$, corresponding to 
$1/3$ of the map and the pixel scale, respectively. 
To analyze the maps and obtain the angular power spectrum we use the software
package HEALPix\footnote{http://www.eso.org/science/healpix/}(G{\'o}rski, Hivon \&
Wandelt 1999). The results
are shown in Fig.~\ref{fig:power} together with the 
primary CMB power spectrum from WMAP data fitting (Spergel et al. 2003).
The signal due to patchy reionization dominates the primary
CMB power spectrum for $l\gsim 4000$ and reaches a maximum amplitude of 
$\approx 1.6\times 10^{-13}$. The amplitude in the two models is comparable.
The power spectrum obtained here is in agreement with that 
derived by Gnedin \& Jaffe (2001), and it is roughly an order of magnitude
smaller than the one calculated by Santos et al. (2003) via a 
semi-analytical model. A last aspect which is worth commenting is the fact 
that the anisotropy keeps a rather flat level up to the highest significant 
multipoles. That is the indication that the secondary anisotropy from reionization 
keep their structure at least up to the arcsecond scale. 

\subsection{Comments on observability}

The detection of the signal from patchy reionization requires high sensitivity
experiments that can reach large multipole numbers, since the peak of the power
spectrum is expected to be at $l$ of the order of few $\times 10^3$.
These characteristics are within the capability of the next generation of 
millimeter wavelength interferometers like 
ALMA\footnote{http://www.alma.nrao.edu/ or www.eso.org/projects/alma}, 
ACT\footnote{http://www.hep.upenn.edu/angelica/act/act.html}, or
CQ\footnote{http://brown.nord.nw.ru/CG/CG.htm}. For example, ALMA is expected to reach sensitivities of 2~$\mu$K rms for a
$1^\prime$ beam with 1~$h$ integration up to 2 arcmin scale,
thus appearing as a perfect instrument to search for
signature of inhomogeneous reionization. 

However, to measure the 
power spectrum from patchy reionization, several other
astrophysical signals must be cleaned out from the maps. In particular, the 
main foregrounds  in the angular range discussed here are the 
thermal Sunyaev-Zel'dovich (SZ) and the Poisson noise from faint point sources.
Thermal SZ is expected to be negligible, at least after multifrequence 
cleaning, for observations at 217 GHz (Zhang, Pen \& Trac 2004). 
More important is the foreground from unresolved IR and radio sources,
which is several orders of magnitude above the reionization signal at
$l\gsim {\rm few}\times 10^4$.
Luckily, this foreground contamination can be described in terms of a
simple power-law (White \& Majumdar 2003). Thus, a foreground measurement at
$l \approx 10^4$ would allow to extrapolate its value at lower $l$, where it
can be subtracted to obtain a clean reionization signal.
For this reason, multifrequency observations are particularly suited to
subtract such foreground contamination.
For this technique to be successful though, a good knowledge of the instrumental
noise is required.

\section{21~cm radiation from neutral IGM}
\label{sec_21cm}

\begin{figure*}
\center{{
\epsfig{figure=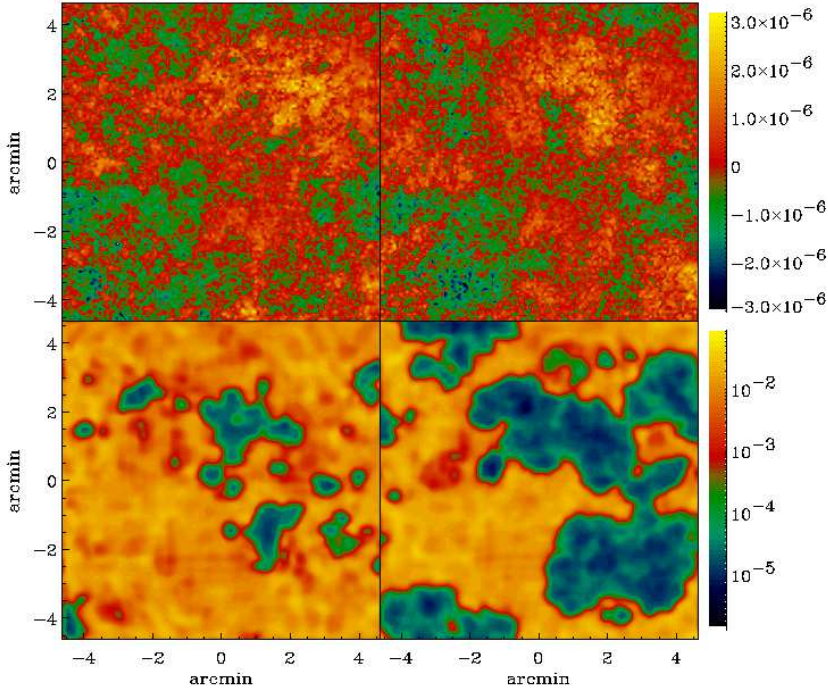,height=10cm}
}}
\caption{{\it Top panels}: map of CMB temperature fluctuations, 
$(\delta T/T)_{\rm CMB}$, due to 
patchy reionization for the S5 (left panel) and the L20 (right panel) model. 
The maps measure $\approx 9.25$~arcmin on a side.
{\it Lower panels:} maps of differential antenna temperature, $(\delta
T/T)_{\rm 21cm}$, from 21cm line emission for the S5 (left panel) and the 
L20 (right panel) model,
at an observed frequency (redshift) $\nu\approx 115$~MHz ($z \approx 11$) and $\nu \approx 90$~MHz
($z\approx 14.5$), respectively. The maps have been derived assuming a bandwidth
$\Delta \nu=1$~MHz and measure $\approx 9.25$~arcmin on a side.} 
\label{fig:maps}
\end{figure*}

The 21cm hyperfine transition of neutral hydrogen in the IGM provides 
a powerful probe to study the era of cosmological reionization. 
In this paper we use the results from the numerical simulation of CM, 
that we briefly describe in this Section. We refer to the above paper for 
further details.

The emission of the 21cm line is governed by the spin temperature,
$T_S$. In the presence of a CMB radiation with $T_{\rm CMB}=2.725
(1+z)$~K, $T_S$ quickly reaches thermal equilibrium with $T_{\rm CMB}$, and a
mechanism is required that decouples the two temperatures. While the
spin-exchange collisions between hydrogen atoms are too inefficient
for typical IGM densities, Ly$\alpha$ pumping
contributes significantly by mixing the hyperfine levels of neutral
hydrogen in its ground state via intermediate transitions to the $2p$
state. If a Ly$\alpha$ background $\simgt 9 \times 10^{-23} (1+z)\,$ergs 
cm$^{-2}$ s$^{-1}$ Hz$^{-1}$ sr$^{-1}$ is present at redshift $z$,
Ly$\alpha$ pumping will efficiently decouple $T_S$ from $T_{\rm CMB}$.
CM find that the diffuse flux of Ly$\alpha$ photons produced by the same
sources responsible for the IGM reionization, satisfies the above
requirement from $z \approx 20$ to the time of complete reionization.  
As the IGM can be easily preheated by
primordial sources of radiation (e.g.~Madau, Meiksin \& Rees 1997;
Chen \& Miralda-Escud\'e 2003), the universe will, most likely, be
observable in 21cm emission at a level that is independent of the
exact value of $T_S$. Variations in the density of neutral hydrogen
(due to either inhomogeneities in the gas density or different
ionized fraction) will appear as fluctuations of the sky brightness
of this transition, and allow, in principle, to map the history of
reionization\footnote{The brightness temperature against the CMB
is defined as $T_b=T_{\rm CMB}{\rm e}^{-\tau}+T_S(1-{\rm e}^{-\tau})$,
where $\tau$ is the optical depth of a patch of IGM in the hyperfine
transition. The differential antenna temperature between the patch
and the CMB is $\delta T_b \simeq (T_S-T_{\rm CMB})\tau/(1+z)$.}.

Using the numerical simulations described in Section~\ref{sec:num}, 
CM have studied the evolution of the 21cm line emission expected
from those reionization histories. In particular, they have derived
maps and fluctuations of brightness temperature at different redshifts 
(i.e. observed frequencies) in both the S5 and L20 model.
The S5 model predicts a peak in the amplitude of the expected rms
brightness fluctuations at an observed frequency (redshift) $\nu\approx
115$~MHz ($z \approx 11$), whereas in the L20 the peak is at 
$\nu \approx 90$~MHz ($z\approx 14.5$). 
In both models, the overall amplitude of the signal at its peak is
$\langle \delta T_{rms}^2 \rangle^{1/2} \approx 10-20$~mK at angular 
scales $\theta \approx 5$ arcmin (see CM for details). 
In Fig.~\ref{fig:maps} maps of differential antenna temperature, $(\delta T
/T)_{\rm 21cm}\equiv \delta
T_b/T_{\rm CMB}(0)$, are shown for the S5 (lower left panel) and the
L20 (lower right panel) model at the peak frequencies. 
The maps have been derived from the simulations of reionization 
described in Section~\ref{sec:num} assuming a bandwidth $\Delta \nu=
\nu \Delta z/(1+z)=1$~MHz and measure $\approx 9.25$~arcmin on a side.

\section{CMB/21~cm Cross-correlation}
\label{sec:cc}

\begin{figure}
\center{{
\epsfig{figure=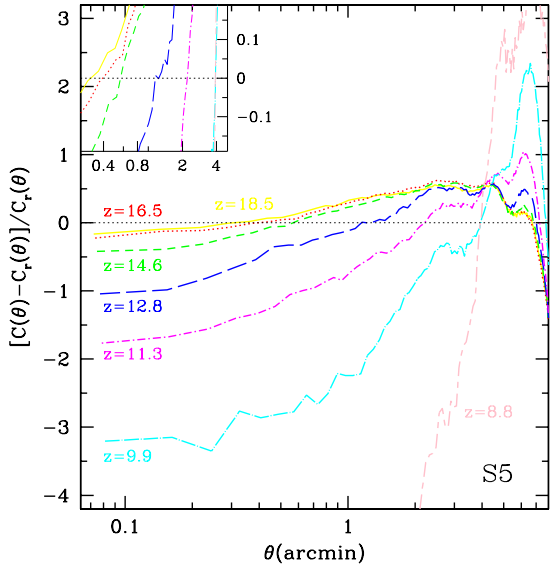,width=8cm}
}}
\caption{Cross-correlation between the CMB secondary anisotropy map and the
21cm maps at different redshifts (marked by different line styles) for
a bandwidth $\Delta \nu=1$~MHz. In the small panel, a zoom of the region 
in which the transition from anti-correlation to correlation occurs is shown. The
subscript {\it r} refers to the random cross-correlation. Model S5. }
\label{fig:ccS5} 
\end{figure}

\begin{figure}
\center{{
\epsfig{figure=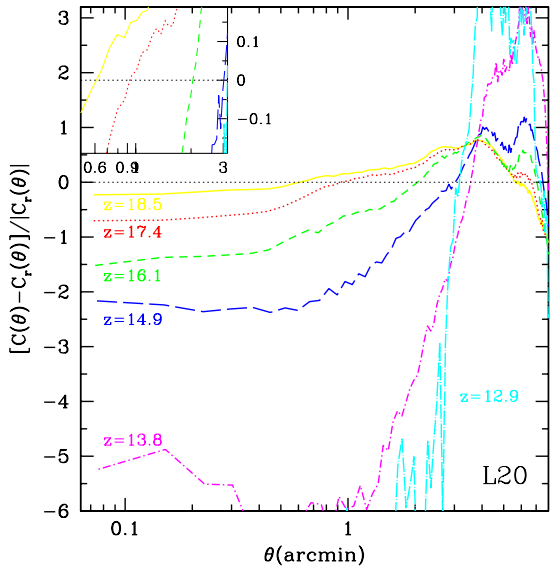,width=8cm}
}}
\caption{The same of Fig.~\ref{fig:ccS5}, but for Model L20.
Note that lines refer to different redshifts.}
\label{fig:ccL20}
\end{figure}

CMB secondary anisotropies from patchy reionization are expected to be
highly anti-correlated with 21cm line emission temperature fluctuations 
on scales smaller than the angle subtended by typical \HII regions 
at the redshift of the 21cm emission.
To quantify this effect, we compute the cross-correlation between 
the CMB and the 21cm map at redshift $z$ as:
\q
C(\theta,z)=\sum_{i=1}^{N_{\rm CMB}}\sum_{j=1}^{N_{\rm 21cm}} \frac{1}{N_\theta}
\left[ \left( \frac{\delta T}{T} \right)_{\rm CMB}(i) 
\left( \frac{\delta T}{T} \right)_{\rm 21cm}(j,z) \right], 
\nq
where $N_{\rm CMB}$ ($N_{\rm 21cm}$) is the number of pixels in the CMB (21cm) map. 
The components of $C(\theta,z)$ are then
binned according to the separation angle $\theta$ between the two lines of 
sight passing through the pixels $i$ and $j$ of the 21cm map at redshift $z$. 
$N_{\theta}$ is the number of values falling in $[\theta,\theta+\theta_{min}]$,
where $\theta_{min}$ is the angular dimension of the cell. We repeat this
procedure for a random binning. 
The results are shown in Fig.~\ref{fig:ccS5} and~\ref{fig:ccL20} for the
S5 and L20 model, respectively.  The subscript {\it r} refers to the random
cross-correlation. The small box in 
the top-left corner shows a zoom of the region in which the majority of the
curves first pass through a zero point, i.e. from anti-correlation to correlation.

As expected, we find that the two signals are highly anti-correlated below
a characteristic angular scale, $\theta_0$, except for the highest redshift where only 
a very small fraction of the volume is ionized (line at $z=18.5$).
The amplitude of the anti-correlation signal increases with decreasing redshift
until reionization is almost completed. At any given 
redshift the model L20 shows a stronger anti-correlation signal, as
reionization proceeds more rapidly, and a larger $\theta_0$,
as a result of the relative larger (on average) \HII region sizes. In fact,
the angular scale $\theta_0$ indicates the typical dimension of the \HII regions 
(including overlapping) at that redshift and allows, in principle, to 
reconstruct the reionization history and to discriminate among different
reionization models and sources (e.g. quasars or massive Pop~III stars versus 
more standard stars).

\begin{figure}
\center{{
\epsfig{figure=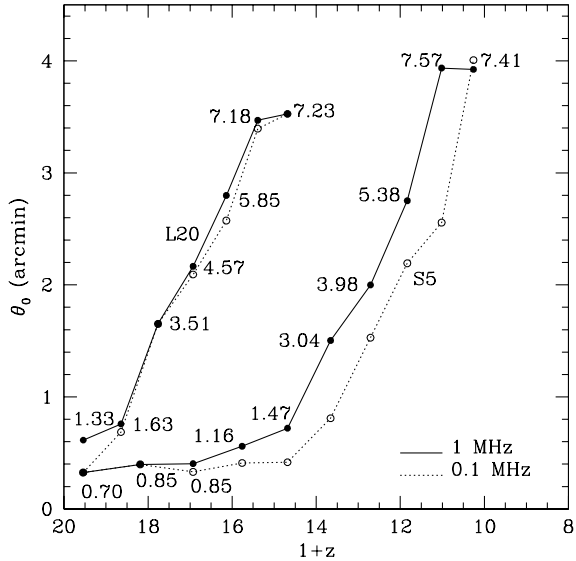,width=8cm}
}}
\caption{Redshift evolution of the angular scale at which the transition 
between anti-correlation and correlation takes place in the L20 (upper curves)
and S5 (lower) models. The labels report
the typical comoving dimension of the corresponding \HII regions (including 
overlapping) in units of Mpc/$h$. The 21cm maps have been obtained assuming 
a bandwidth $\Delta \nu=1$~MHz (solid lines) and~0.1~MHz (dotted).}
\label{fig:theta_z}
\end{figure}

The redshift evolution of the angular scale $\theta_0$ is shown in 
Fig.~\ref{fig:theta_z}; labels there indicate the typical
comoving dimension of the corresponding \HII regions in units of Mpc/$h$. 
As the evolution of $\theta_0$ reflects the growth of \HII regions,
the value of $\theta_0$ at a given redshift is very different for the two
models considered. In general, this result can be used to discriminate among 
different ionizing sources as \HII regions produced by quasars 
or massive Pop~III stars typically tend to be larger than those digged by more standard stars.
Moreover, as the redshift
evolution of $\theta_0$ reflects the growth of \HII regions, 
non-monotonic reionization histories would result in a more complex behavior for $\theta_0$.
For example, in case of a double-reionization (e.g. Cen 2003; Wyithe \& Loeb 2003) 
we expect that $\theta_0$ increases until the first reionization
is completed. Then, once the ionizing emissivity drops and the IGM partially
recombines, $\theta_0$ should decrease or remain constant, and eventually grow again 
when the second reionization takes place.

\begin{figure}
\center{{
\epsfig{figure=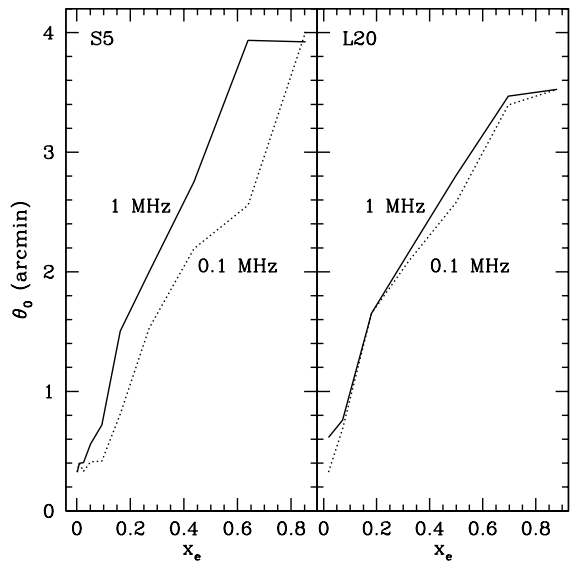,width=8cm}
}}
\caption{Relation between the zero-point angle of the cross-correlation function  
and the mean ionization fraction in the simulation box
for the S5 (left panel) and L20 (right) models. 
The 21cm maps have been obtained assuming 
a bandwidth $\Delta \nu=1$~MHz (solid lines) and~0.1~MHz (dotted).}
\label{fig:theta_xe}
\end{figure}

From our simulations it is also possible to derive a relation between the measured
value of $\theta_0$ and the volume averaged hydrogen ionization fraction in 
the computational volume. This relation is shown in Fig. \ref{fig:theta_xe}.  
From there we see that
the two quantities are positively correlated. It is also worth noticing that a
50\% mean ionization level is characterized by $\theta_0 \approx 3'$ independently 
of the adopted reionization model: 
identifying this epoch is crucial as it corresponds
to the redshift at which most of the CMB secondary anisotropies are produced and it provides
a sensible definition of the reionization redshift for prompt reionization models often adopted 
for practical purposes (Bruscoli, Ferrara \& Scannapieco 2002).
Moreover, the $\theta_0$ - $x_e$ relation appears to 
be quite insensitive to the details of the reionization model, thus providing a
robust mapping between the correlation function and the mean ionization level 
at each epoch.

The results in Fig. \ref{fig:ccS5} and \ref{fig:ccL20} have been derived assuming a bandwidth for 21cm line observations
of $\Delta \nu=1$~MHz. Although a smaller bandwidth, e.g. 0.1~MHz, would substantially
increase the intensity of 21cm line emission (CM), it does not significantly
affect the estimates of the cross-correlation (Fig.~\ref{fig:theta_z}). 

In conclusion, we find that the cross-correlation between secondary anisotropy 
in the CMB and 21cm emission maps can be a useful tool to follow the 
reionization process and to give constraints on the nature of the ionizing 
sources. Moreover, the cross-correlation, combining information obtained by
different experiments, can be used to maximize the signal from the 
reionization process with respect to instrumental noise, systematic errors in 
the measures, and astrophysical foregrounds. 

\section{Summary}      
\label{sec:concl}

We have calculated the secondary anisotropies in the CMB temperature power 
spectrum produced by inhomogeneous reionization from radiative transfer
simulations consistent with WMAP observations.
We find that the signal arising from this process dominates
over the primary CMB component for $l\gsim 4000$ and reaches a 
maximum amplitude of $l(l+1)C_l/2\pi\simeq 1.6\times 10^{-13}$ on arcminute 
angular scales, i.e. $l$ as large as several thousands.
We then cross-correlated the secondary CMB anisotropy maps with 
21cm line emission fluctuation maps for the same reionization simulations.
As expected, the two signals are highly anti-correlated on 
angular scales corresponding to the typical size      of \HII regions 
(including overlapping) at the redshift of the 21cm map. The cross-correlation
and, in particular, the redshift evolution of the angular scale at which the
transition between anti-correlation to correlation takes place, can be used:
(a) to study the nature of the reionization sources, (b)  
to reconstruct the cosmic reionization history, (c) to infer the mean cosmic   
ionization level at any redshift.

Cooray (2004) has studied the correlation signal between CMB temperature
anisotropies and 21cm fluctuations by means of an analytical
reionization model. He concludes that, contrary to what we have shown, 
the correlation cannot be seen in the angular cross-power spectrum, due
to a geometric cancellation effect between velocity and density 
fluctuations. However, such cancellation likely occurs
due to the assumption made in that paper that the neutral hydrogen
fraction depends only on overdensity but not on spatial location. 
In fact, Cooray writes the
neutral hydrogen density as $n_{HI}=x_H \bar n (1+\delta)$, where $x_H$
is the neutral H fraction, $\bar n$ is the mean gas density and $\delta$
is the gas overdensity. In a patchy reionization scenario, where the
ionized bubbles around luminous sources do not completely fill the cosmic volume,
it is clear that this assumption is not correct, as two
fluctuations with the same $\delta$ value located either inside a
ionized region or outside it will have different $n_{HI}$. Hence,
if the patchiness of the reionization can be properly modelled (\eg through
radiative transfer simulations) the degeneracy (and the above
cancellation) can be broken. Notice that the largest contribution to 
the secondary anisotropies come from the epoch when roughly 50\% of the cosmic
volume is filled with bubbles and where the variance in the relation
$n_{HI} - \delta$ is largest. 

Planned millimeter wavelength interferometers, like ALMA and ACT, are expected to
have sensitivities and angular resolution good enough to measure the signature 
of inhomogeneous reionization in the CMB maps. 
However, extracting information on the reionization process from the observed maps
can be hampered by the presence of both astrophysical foregrounds and 
instrumental noise. The same applies to 21cm emission observations 
(see Di Matteo, Ciardi \& Miniati 2004 for a detailed study of the foreground 
contamination of 21cm maps). Provided that both the CMB and 21cm maps can be cleaned
from foreground contamination, the information obtained from a cross-correlation
of the two maps is an invaluable tool to study the reionization history and its sources.

\section*{Acknowledgments}

Some of the results in this paper have been derived using the HEALPix (G{\'o}rski, Hivon, and Wandelt 1999) package.

\end{document}